\documentclass{article}
\usepackage{arydshln,fontenc, inputenc, calc, indentfirst, fancyhdr, graphicx,epstopdf, lastpage, ifthen, lineno, float, amsmath, setspace, enumitem, mathpazo, booktabs, titlesec, etoolbox, tabto, xcolor, soul, multirow, microtype, tikz, totcount, amsthm, bm, footmisc, url, newfloat, caption}
\usepackage[a4paper, total={6.5in, 10in}]{geometry}
\usepackage[hyperfootnotes=false]{hyperref}
\usepackage[authoryear, round]{natbib}
\begin{document}
\title{Statistical Oceanography of Profiling Floats and Surface Drifters}
\author{Mikael Kuusela\footnote{ Department of Statistics and Data Science, Carnegie Mellon University, Pittsburgh, USA; email: mkuusela@andrew.cmu.edu},\, Sofia C. Olhede\footnote{Institute of Mathematics, École Polytechnique Fédérale de Lausanne, Lausanne, Switzerland; email: sofia.olhede@epfl.ch} \, and Adam M. Sykulski\footnote{Department of Mathematics, Imperial College London, London, UK; email: adam.sykulski@imperial.ac.uk}}
\date{}
\maketitle
\begin{abstract}
The Global Ocean Observing System is key to understanding oceanic variability and climate change. The ocean is vast and often sparsely and irregularly sampled in time and space by various instruments and systems. Statistical models, especially spatio-temporal ones, are useful for enabling inferences, forecasts and decisions from sparse oceanic observations. This article focuses on the statistical treatment of two types of in situ observations in the Global Ocean Observing System: from profiling floats and surface drifters, focusing on the Argo and Global Drifter Programs. We describe the spatio-temporal models that have been developed in recent years for these data, to give a picture of the statistical challenges faced in modern physical oceanography. We will discuss in detail two types of reference frames when modeling such data: Eulerian and Lagrangian, the former of which is more appropriate for profiling floats and the latter for surface drifters.
\end{abstract}
\vspace{3mm}
{\em This article has been accepted for publication in the Annual Review of Statistics and Its Application, Volume 14.}
%
\section{Introduction}
%

The Global Ocean Observing System (GOOS) is crucial to monitoring and understanding the role of ocean variability in the Earth system; see, for example,~\cite{legler2015current,miloslavich2019challenges,Wong2020} or \cite{elipot2022dataset}. Because the ocean occupies roughly 71\% of the Earth's surface, in situ monitoring of the ocean with both uniform sampling and good coverage in space and time is very challenging. Numerous data sources feed into the GOOS, but in this article we focus on {\em profiling floats} (in particular the Argo Program, a global array of profiling floats that measure temperature, salinity, pressure and other variables at different depths) and {\em surface drifters} (in particular the Global Drifter Program of satellite-tracked drifting buoys used to estimate near-surface currents, temperature, and other variables at the surface). Analysis of these massive datasets involves statistical challenges ranging from irregular and non-standard sampling, to the modeling of complex non-stationary, non-Gaussian and multivariate spatio-temporal stochastic processes and random fields. The importance of the Argo and Global Drifter Programs comes from how complementary and uniquely scalable they are within the GOOS. They provide sustained, global, in situ measurements that no other platform can match in their combination of coverage and resolution. Together they allow oceanographers to jointly resolve the three-dimensional, time-evolving state of the global ocean in a way that no other observing system can.

Data from profiling floats and surface drifters provide an ideal breeding ground for new statistical methodology. These data have complex spatio-temporal structure that pushes the boundaries of standard inference techniques. The datasets are large enough to fit sophisticated models, but sparse enough that methodological choices have non-trivial impacts on the final scientific inferences. This `Goldilocks' setting has motivated a flurry of recent statistical work, which we review here with the goal of introducing the broader statistical community to these rich datasets and the associated opportunities. For a broader methodological review of statistical techniques across oceanography, see \cite{wikle2013modern, wikle2019statistics} and \cite{vanem2022statistical}.

With profiling floats and surface drifters, the choice of methodology will vary by instrument and goal, and a key separation is into Eulerian observations (made in a fixed reference frame) and Lagrangian observations (made in a moving reference frame). We also discuss the peculiarities of both observational platforms, the motivating science questions, statistical modeling of these observations, and finally how advances in data integration and machine learning aim to combine the information obtained from float and drifter data with satellite remote sensing and other instruments.

This article starts in Section~\ref{sec:measure} with background on the instruments and data, the broader scientific motivation, and a discussion of the reference frames. In Section~\ref{ss:floats}, we discuss statistical analysis techniques for profiling float data, and in Section~\ref{ss:drifters} for surface drifter data. Section~\ref{sec:dataintegration} discusses data integration using techniques such as data fusion and data assimilation. We conclude with a discussion and future outlook in Section~\ref{sec:discandfuture}.

\section{Background}\label{sec:measure}
\subsection{Scientific Motivation}
Physical oceanography \citep{talley2011descriptive} studies the ocean's physical properties, such as ocean circulation, ocean-atmosphere interactions, variability in temperature and salinity, and the impact of these processes on the global climate system. Any understanding of climate fundamentally depends on our understanding of the ocean, which plays a critical role in redistributing and storing heat as well as in the global hydrological, carbon and oxygen cycles. For this reason, there have been numerous efforts to monitor key ocean properties over space and time with various global and regional observational platforms. Measuring the properties of the ocean is not simple: the ocean covers 71\% of the Earth's surface, including extremely remote locations, and has an average depth close to 4000 m. 

Among the measurement programs introduced to understand the ocean, this article is mainly focused on floats and drifters, but we also mention ship-based observations, moored instruments, such as the tropical Pacific TAO/TRITON array and the PIRATA array in the Atlantic, and satellites which provide remotely sensed observations of sea surface height (SSH), sea surface temperature (SST), ocean color and other variables. Together, these and other instruments form the GOOS. These platforms can broadly be categorized according to whether they provide measurements at the ocean surface (drifters and satellites) or subsurface (floats, ships and moorings). Another important distinction is between remotely sensed (satellite) and in situ observations (the other platforms mentioned).

A key observational and statistical challenge is that the ocean is very inhomogeneous. It is strongly stratified in the vertical dimension, so water properties vary at different depths. Additionally, many variables tend to change rapidly meridionally (going north to south), but less rapidly zonally (going east to west). This complicates statistical modeling, as stationary and spatially isotropic methods will not, in general, perform well.

\subsection{Introduction to Profiling Floats and Surface Drifters} \label{sec:instruments}

The easiest way to track surface currents is to deploy an instrument that floats at the surface and moves with the currents, and then track it using satellites. Such instruments are called \emph{surface drifters}. On the other hand, \emph{profiling floats} dive below the surface and give information about subsurface currents and other oceanographic variables such as subsurface temperature and salinity. It is perhaps confusing that the surface instruments float and are called drifters, while the subsurface instruments dive and are called floats.

\begin{figure}
\includegraphics[width=10.7cm]{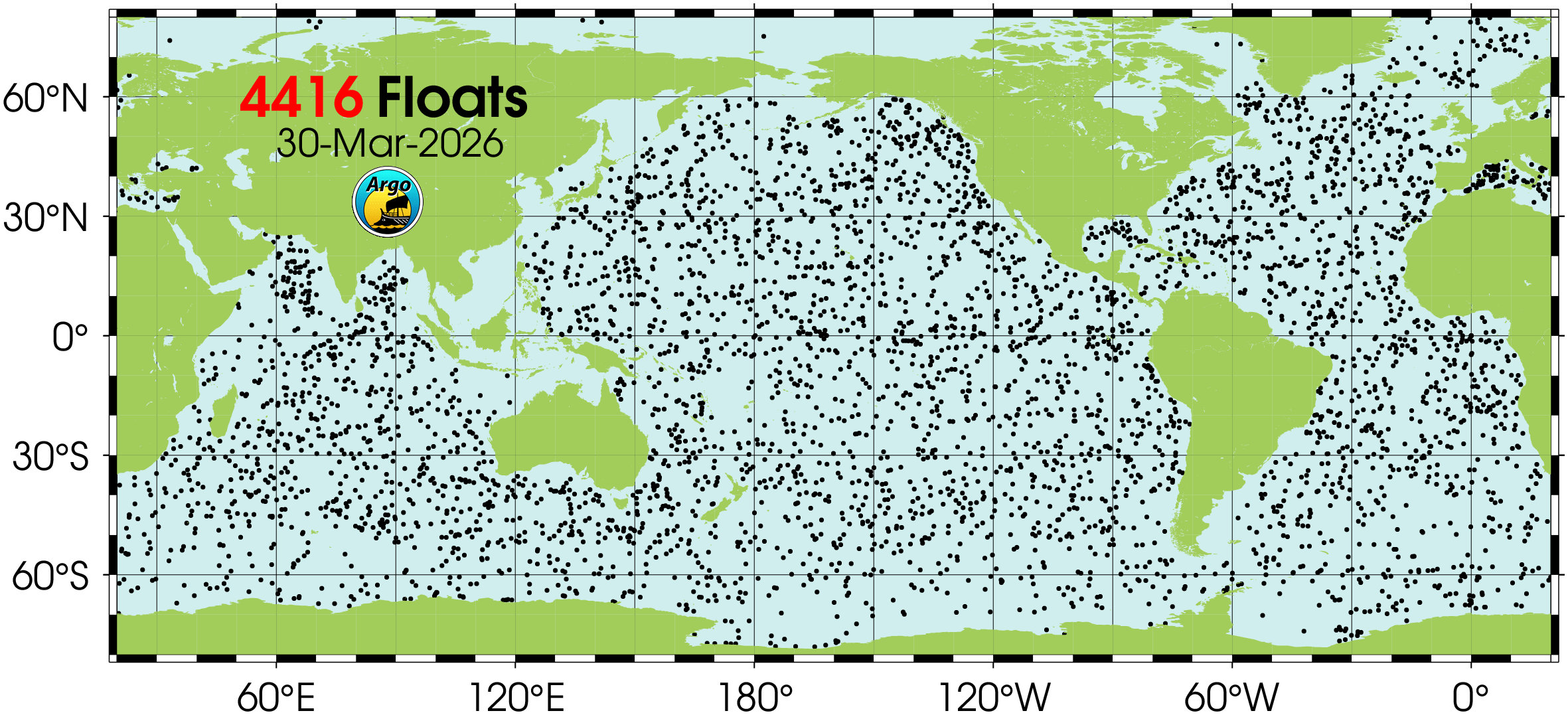}
\hfill
\includegraphics[width=5.5cm]{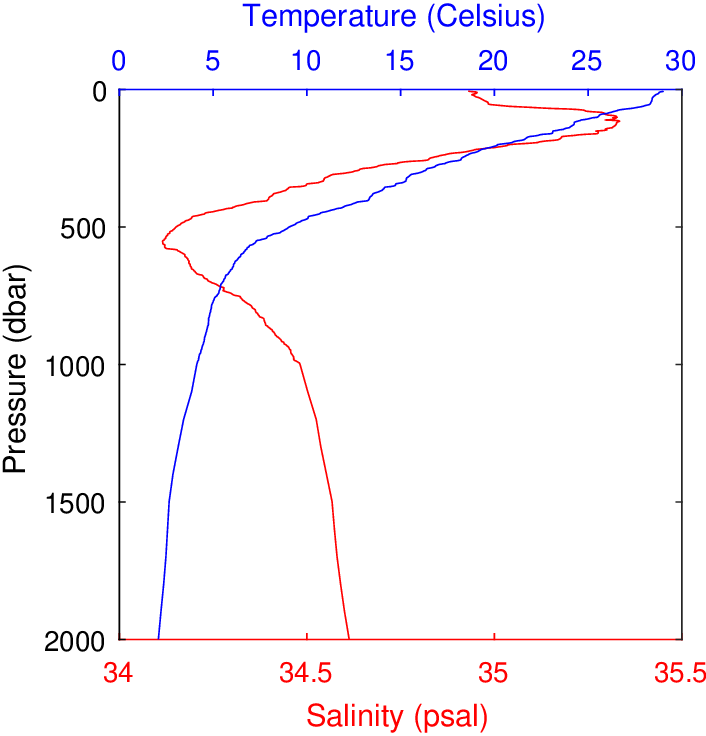}
\caption{Illustration of Argo float data. (left) Argo float locations on March 30, 2026. (Figure credit: The Argo Program, \url{https://argo.ucsd.edu/about/status/}) (right) Example Argo temperature (T) and salinity (S) profiles. These particular profiles were collected on August~30, 2015 in the Western Pacific Ocean (20.0$^\circ$N 148.1$^\circ$E).}
\label{fig:argo}
\end{figure}

The Argo array \citep{Roemmich2009_ArgoProgram, Riser2016, Wong2020} is the most comprehensive float-based subsurface observational program. It consists of approximately 4,000 autonomous floats that sample the upper 2,000 m of the ocean globally at roughly $3^\circ \times 3^\circ$ spatial resolution. Figure~\ref{fig:argo} (left) shows the locations of the Argo floats on March~30, 2026. Each float follows a regular, in most cases 10-day, cycle, which starts when it descends to its parking depth at 1,000 m. After drifting for roughly 10 days, it further descends to 2,000 m. From there, it starts an ascent to the surface, during which it collects vertical profiles of temperature and salinity as a function of pressure, as shown in Figure~\ref{fig:argo} (right). Once the float reaches the surface, it records its position using satellite positioning, transmits the data to ground stations via satellite and starts the next profiling cycle.

Argo floats have been deployed by various countries since 1999. Sparse global coverage was reached around 2004 and the target $3^\circ \times 3^\circ$ sampling resolution in 2007. To date, the Argo array has collected more than 3 million profiles. As a result, Argo data provide a unique, more-than-20-year, seasonally and spatially uniform record of the global evolution of subsurface ocean temperature, salinity and currents. In recent years, the array has been expanded in two ways \citep{jayne2017argo}: the Deep Argo Program \citep{zilberman2023observing} develops and deploys floats that sample down to 6,000 m depth, providing observations of the rarely sampled abyssal ocean. The Biogeochemical-Argo (BGC-Argo) program \citep{Claustre2020} adds new sensors to the floats to measure oxygen, nitrate, pH, chlorophyll fluorescence, optical backscatter and downwelling irradiance, providing subsurface information on ocean biogeochemistry. Since Deep Argo and BGC-Argo floats are more expensive than standard Core Argo floats, the current design of the Argo array calls for a mix of the three different float types to be used.

\begin{figure}
\centering
\includegraphics[width=0.87\textwidth]{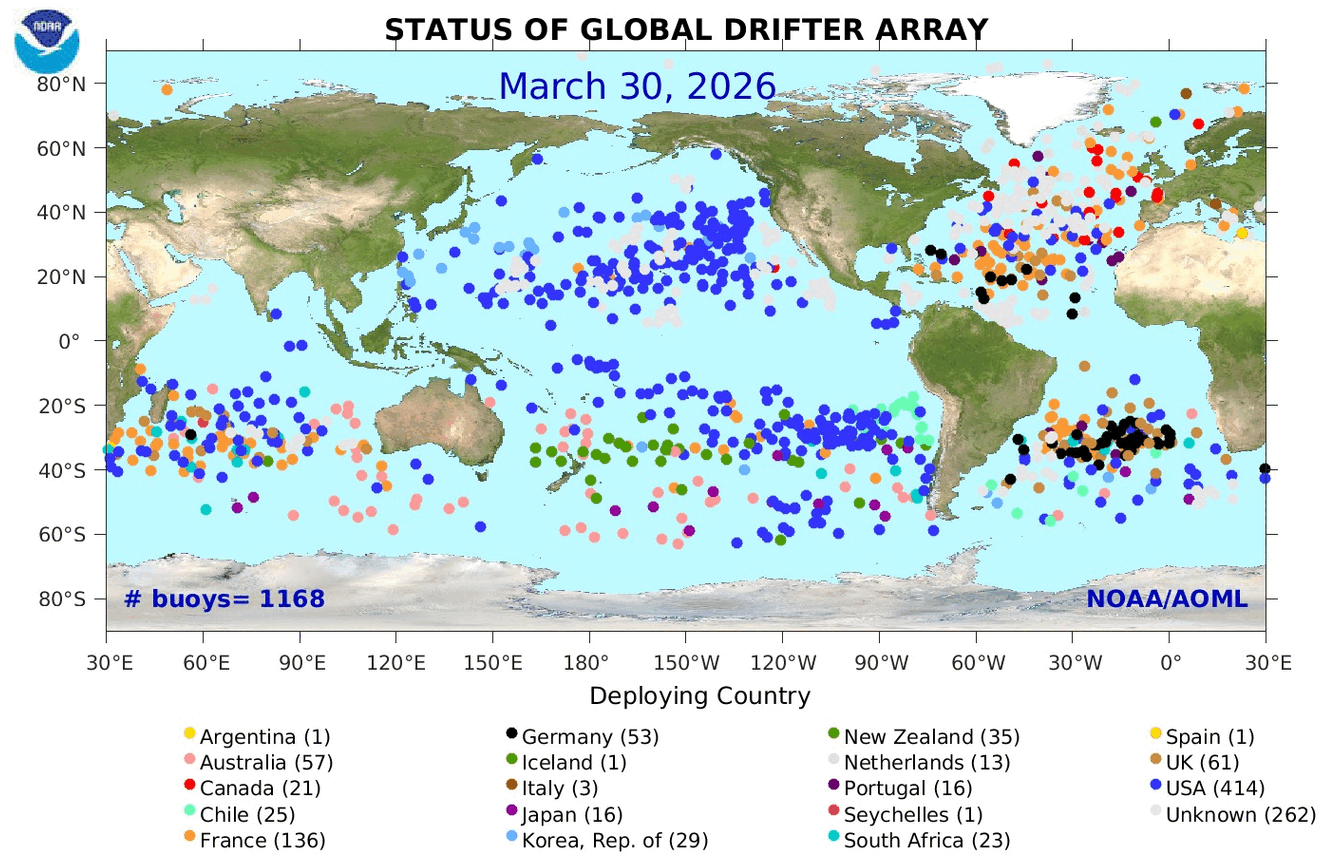}
\caption{Locations of the Global Drifter Program buoys on March 30, 2026, with breakdown by deploying country. (Figure credit: Atlantic Oceanographic and Meteorological Laboratory (NOAA), \url{https://www.aoml.noaa.gov/phod/graphics/dacdata/globpop_countries.gif}.)}
\label{fig:locations}
\end{figure}

Surface drifters, on the other hand, are buoys that freely float at or near the ocean surface, and are transported by currents and surface winds. They are designed to mimic the motion of a particle of water near the surface, so that tracking a drifter's location can be considered as a proxy for directly measuring ocean currents in a Lagrangian (moving) reference frame (see Section~\ref{ss:frame}). The most comprehensive drifter-based observational program is the Global Drifter Program \citep{lumpkin2017advances} managed by NOAA (the US National Oceanic and Atmospheric Administration), which started in 1979 and since 2005 has maintained a network of 1,000--1,500 satellite-tracked drifting buoys distributed across the global ocean on a $5^\circ \times 5^\circ$ grid. It was the first component of the GOOS to reach completion. Figure~\ref{fig:locations} shows the locations of the drifters in the array on March 30, 2026. Each year about 1,000 new drifters are deployed by international partners to replace units that fail or drift out of coverage, and over 30,000 drifters have been deployed in total. Individual drifters are battery-powered and typically designed to last at least 18 months, although exceptional units have transmitted data for over 10 years. The drifters continuously measure variables such as surface current velocities, SST, salinity, sea level and atmospheric pressure, wave properties, and biogeochemical variables such as chlorophyll fluorescence, transmitting near-real-time data that are critical for weather forecasting, climate research, and ocean circulation studies. We shall restrict our review to drifter locations and surface current velocities, as these are considered the core variables of the Global Drifter Program. To ensure that they accurately follow near-surface ocean currents rather than being pushed directly by wind, most drifters are equipped with a subsurface drogue (sea anchor) centered at about 15 m depth, which stabilizes their motion and improves data quality. 

\begin{figure}
\centering
\includegraphics[width=0.87\textwidth]{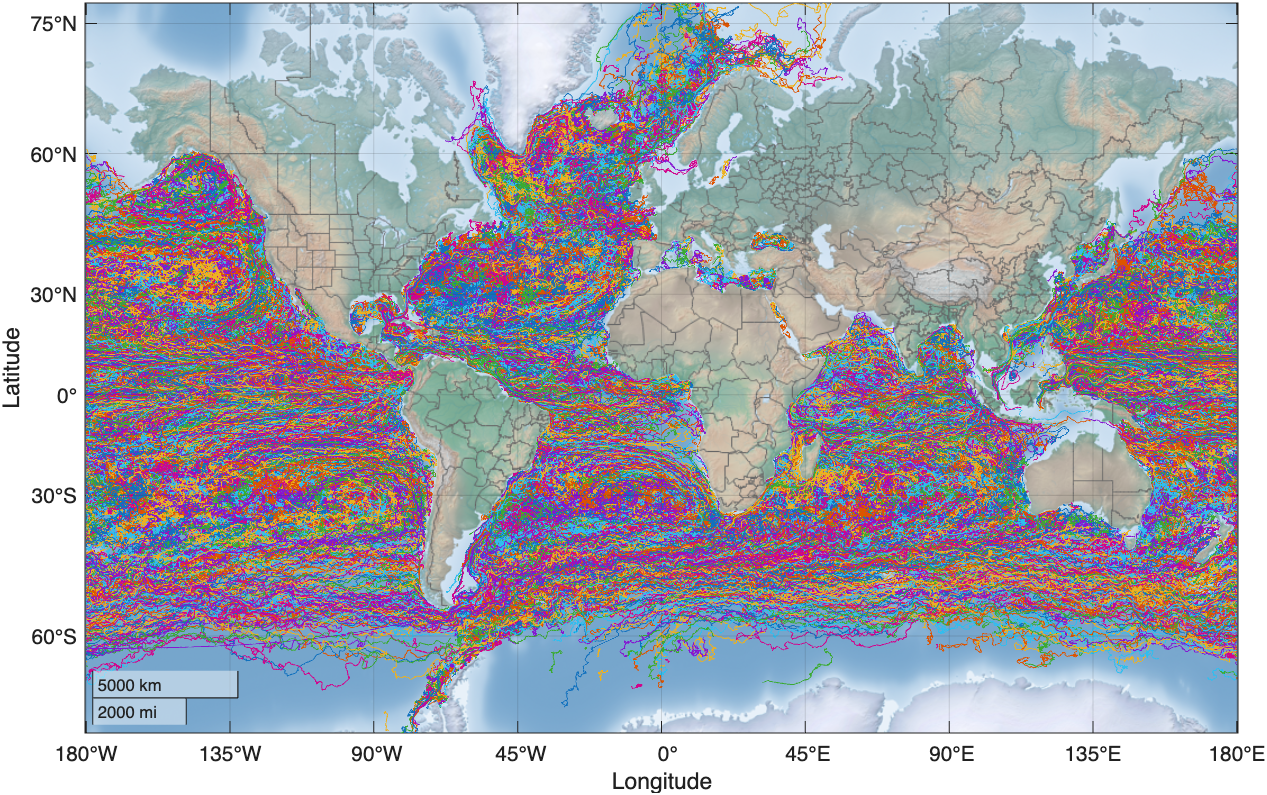}
\caption{A `spaghetti' plot of 10,000 drifter trajectories in the Global Drifter Program database. Each strand of spaghetti represents a single satellite-tracked trajectory. Data available at \url{https://www.aoml.noaa.gov/phod/gdp/data.php}.}
\label{fig:spaghetti}
\end{figure}

Drifter locations are observed as trajectories, as shown in the `spaghetti plot’ in Figure~\ref{fig:spaghetti}. 
Drifters are buoys that move freely with ocean currents, whereas moorings
 are buoys anchored to the seafloor or are otherwise fixed and thus do not
drift. The advantage of drifters (and profiling floats) is their broad spatial coverage, as
moorings are costly and difficult to deploy and maintain, particularly in
deep or remote regions of the ocean.

Argo and Global Drifter Program data are primarily transmitted via the Iridium satellite constellation, typically using GPS for positioning. Previously, positioning and data transmission of both programs mainly used the Argos satellite system (not to be confused with Argo), which limited position accuracy and the quantity of data transmitted. As a result, there is heterogeneity in the quality and quantity of data between the early Argos and modern Iridium eras. Argos remains in use at time of writing, but in a limited capacity.

\subsection{Data Availability}

Argo data are freely available within 24 hrs of data collection via two Global Data Assembly Centers (GDACs), one in France and another in the US. Argo GDAC data are publicly accessible online through various protocols, with up-to-date access information available at \url{https://www.argodatamgt.org/}. Monthly snapshots of the Argo GDACs are automatically produced and accessible through the citable DOI \url{https://doi.org/10.17882/42182}. Using these monthly snapshots is the recommended way to access the raw Argo data. However, these raw data must be filtered before use with various quality-control flags documented in \citet{argo_guide2025}. The \texttt{argopy} Python package \citep{Maze2020}, \texttt{argoFloats} R package \citep{kelley2021argofloats} and Argovis API \citep{mills2025argovis} facilitate access to and preprocessing of Argo data. In addition, a number of curated quality-controlled vertical-profile collections, which include Argo data but usually also other data sources, are available; examples include the EN4 database \citep{good2013en4} and the World Ocean Database \citep{mishonov2024world}.

Global Drifter Program data are freely available at \url{https://www.aoml.noaa.gov/phod/gdp/data.php} where interpolated time series of over 30,000 drifters are available at hourly resolution \citep{elipot2016global}, yielding hundreds of millions of observations in total. There are also several regional surface drifter deployment studies, including GLAD and LASER in the Gulf of Mexico \citep{poje2014submesoscale,miron2017lagrangian}, and subsurface drifter studies using acoustic rather than GPS tracking \citep{veneziani2004oceanic}. Many of the statistical methods reviewed in this paper 
have been, or could be, applied to these regional studies as well---although the specification of the drifters varies by program, so data between programs should be combined with care. Another option for accessing and analyzing data from the Global Drifter Program, as well as various regional drifter deployment programs, is via the \texttt{clouddrift} Python package \citep{elipot2024clouddrift}.

\subsection{Choice of Frame of Reference}
\label{ss:frame}
In oceanography, and fluid dynamics more broadly, two main observational frameworks are commonly distinguished: measurements made while moving with the flow (following a parcel of fluid in space and time), and measurements made at a fixed location (recording how fluid properties vary at a point over time). The former is referred to as a Lagrangian framework, and the latter as an Eulerian framework~\citep{lacasce2005eulerian,lacasce2008statistics}.
An analogy is to consider two ways of tracking the progress of runners in a marathon. One approach is to follow individual runners—for example using wearable devices—to record their trajectories and speeds (a Lagrangian perspective). The other is to record the times at which runners pass fixed checkpoints along the course (an Eulerian perspective).

Given the vast area of the ocean surface, as well as the ocean at depth, it is impossible to acquire global in situ observations at fixed locations in a regularly sampled fashion. In fact the two major measurement programs we discuss in this paper are both based on using forms of drifting (Lagrangian) instruments. Despite this, Eulerian methods are typically used for analyzing float data (Section~\ref{ss:floats}), while Lagrangian methods are more common for drifter data (Section~\ref{ss:drifters}), the motivation for which shall become apparent. 

Within the Eulerian perspective, several choices of the coordinate system in the fixed frame of reference are possible. An obvious choice is to describe the four-dimensional ocean in coordinates of longitude $x$, latitude $y$, depth or pressure $z$ (we use depth and pressure interchangeably since $1 \text{ m} \approx 10 \text{ dbar}$ throughout the water column) and time $t$. Most of the methods we review in Section~\ref{ss:floats} use this coordinate system. However, since the ocean is vertically stratified so that water density increases with depth, an alternative coordinate system sometimes used replaces depth $z$ by density $\rho$ \citep[e.g.,][]{schmidtko2013mimoc}. Due to the physics of ocean dynamics, oceanographic fields tend to have more homogeneous structure on fixed density levels (isopycnals) than on fixed pressure levels (isobars). As such, it might be easier to write down statistical models on isopycnals than on isobars.

Within the Lagrangian perspective, a drifter moves on the ocean surface over time, which we can model using spherical coordinates $\{\lambda(t),\phi(t)\}$. The velocities of the drifter can then be modeled using $\{u(t),v(t)\}$, where $u(t)$ is known as the {\em zonal} velocity (the speed along lines of equal latitude, i.e., east-west) and $v(t)$ is known as the {\em meridional} velocity (the speed along lines of equal longitude, i.e., north-south), typically given in SI units of meters per second. In practice, we may also choose to convert the drifter positions $\{\lambda(t),\phi(t)\}$ to displacements $\{x(t),y(t)\}$ in SI units of meters, using a suitable map projection. 

Any global statistical model for ocean data needs to account for the curvature of the Earth. Writing down valid stochastic models on a sphere is not trivial (see \citet{jeong2017spherical} and the references therein). Some proposed solutions for float data are given in \citet{baugh2022hierarchical} and \citet{salvana20223d} and involve using a careful choice of distance metric. An alternative is to use local spatial models \citep{kuusela2018locally} which approximate small regions of the spherical surface as Euclidean.

\section{Statistical analysis of profiling float data}\label{ss:floats}

Argo floats are designed to observe four-dimensional oceanographic fields on a global scale. Let $f(x,y,z,t)$ be some oceanographic field of interest (such as temperature, salinity or density) indexed by longitude $x$, latitude $y$, depth $z$ and time $t$. As described in Section~\ref{sec:instruments}, the basic units of data produced by Argo floats are vertical profiles $\{f(x_i,y_i,z_{ij},t_i)\}_{j=1}^{n_i}$ through the field $f$, where $(x_i,y_i,t_i)$ is the spatio-temporal location and $z_{ij},\,j=1,\ldots,n_i$, are the sampling depths of the $i$th profile. Core and Deep Argo floats directly observe profiles of the temperature and salinity fields; BGC-Argo floats additionally observe some subset of six biogeochemical variables. Further fields, such as density, which is a nonlinear function of temperature, salinity and pressure, can be derived from these primary observations.

While Argo data have motivated a wide range of statistical research, ranging from estimating tropical cyclone signals \citep{cheng2015global,hu2024spatiotemporal} to trajectory inference \citep{chamberlain2018observing,chamberlain2023using,hansen2022probabilistic}, the vast majority of work on these data has focused on the spatio-temporal mapping problem of statistically interpolating the observations onto a regular four-dimensional grid across longitude, latitude, depth and time. Interpolation across the vertical dimension within a profile is relatively straightforward since this dimension is more densely observed than the others. A standard approach is to use piecewise cubic Hermite interpolating polynomials (PCHIP) to interpolate the vertical observations into a continuously resolved vertical profile. Thanks to their shape-preserving property, PCHIP have been shown to provide highly accurate interpolations even for the more sparsely sampled early Argo profiles \citep{Barker2020, Li2022, sukianto2026locally}. Alternative approaches \citep{yarger2022functional,korte2025functional} use functional data analysis, where the interpolation is done simultaneously across all four dimensions and the vertical dimension is handled using a basis expansion.

Interpolation across longitude, latitude and time is statistically more challenging. The approach taken in most mapping approaches either explicitly or implicitly considers a random field model of the type
\begin{equation}
f(x,y,z,t) = \mu (x,y,z,t) + a(x,y,z,t) + \varepsilon(x,y,z,t),
\label{eq:basic_model}
\end{equation}
where $\mu (x,y,z,t)$ is a deterministic mean field (also called the \emph{climatology} in oceanography and climate science), $a(x,y,z,t)$ is a spatially and temporally correlated zero-mean random field (also called the \emph{anomaly} field) and $\varepsilon(x,y,z,t)$ is a nugget effect that is uncorrelated across $x$, $y$ and $t$, but could be correlated in $z$. In this model, $\mu (x,y,z,t)$ reflects the long-term climatological mean state of the ocean and $a(x,y,z,t)$ reflects spatially and temporally resolved transient deviations from this mean state. The nugget effect $\varepsilon(x,y,z,t)$ corresponds to unresolved small-scale variation, capturing both oceanographic variation at scales smaller than the Argo sampling resolution (such as from eddies and other mesoscale phenomena) and variation due to instrument noise. The importance of instrument noise varies by sensor: for the Argo temperature observations, instrument noise is negligible for most purposes, but it can be large for the BGC and salinity sensors in particular.

The spatio-temporal interpolation task is then to use the model of Equation~\ref{eq:basic_model} to produce predictions $\hat{f}(x^*,y^*,z^*,t^*)$ of the field on a regular grid of unobserved locations $(x^*,y^*,z^*,t^*)$. The end result is a \emph{gridded data product} that is used by scientists in various downstream studies. Many such Argo data products have been produced by the oceanographic community; examples include the Roemmich--Gilson product \citep{Roemmich2009}, EN4 \citep{good2013en4}, MIMOC \citep{schmidtko2013mimoc} and ISAS \citep{gaillard2016situ}. The rest of this section provides an overview of the models and methods developed in recent years by the statistical community to produce improved Argo gridded products.

\subsection{Modeling non-stationarity}

One of the key challenges in writing down a concrete form for the model in Equation~\ref{eq:basic_model} is the heterogeneous structure and variation of most oceanographic fields. Statistically, this means that the random field $f(x,y,z,t)$ is strongly non-stationary. The most obvious form of non-stationarity is in the mean field $\mu(x,y,z,t)$, which varies by longitude, latitude, depth and season. Additionally, due to long-term climate change in the ocean, it can be reasonable to include a long-term time trend in $\mu(x,y,z,t)$ \citep{sukianto2026locally, baugh2022hierarchical}.

For most oceanographic fields, $\mu(x,y,z,t)$ has rich structure across the spatial coordinates. As a result, it is hard to write down a global parametric model for $\mu(x,y,z,t)$, so a nonparametric approach is usually preferred. A popular nonparametric model for $\mu(x,y,z,t)$ is to use local regression \citep{Ridgway2002,Roemmich2009}. To obtain an estimate of $\mu$ at $(x^*,y^*,z^*,t^*)$, one pools data across time that is spatially close to $(x^*,y^*,z^*)$ and fits a regression function locally in space to these data. \citet{Ridgway2002} and \citet{Roemmich2009} considered the regression function
\begin{equation}
\begin{split}
\mu(x,y,z,t) =&\beta_0+\beta_1(x-x^*)+\beta_2(y-y^*)\\
&+\beta_3(x-x^*)(y-y^*)+\beta_4(x-x^*)^2+\beta_5(y-y^*)^2\\
&+\beta_6 (z-z^*) + \beta_7 (z-z^*)^2 \\
&+\sum_{k=1}^6\gamma_k\sin\left(2\pi k \frac{\tau(t)}{365}\right)+\sum_{k=1}^6\delta_k\cos\left(2\pi k \frac{\tau(t)}{365}\right),
\end{split}
\label{eq:mean_field}
\end{equation}
where $\tau(t) \in [0,365]$ is the yearday corresponding to time $t$. Here the first three lines fit a quadratic surface in $(x,y,z)$ and the last line uses harmonics with periods of 2--12 months to model the seasonal cycle. The regression is typically fitted using ordinary least squares and may include weights to downweight observations further away. This ignores the spatio-temporal dependence in $a(x,y,z,t)$, although methods to jointly fit $\mu$ and $a$ have also been developed \citep{gray2015method,baugh2022hierarchical, park2023spatiotemporal}. The estimated mean field at $(x^*,y^*,z^*,t^*)$ is then the fitted value of this regression $\hat{\mu}(x^*,y^*,z^*,t^*) = \beta_0 + [\text{seasonal harmonics at } t^*]$ and the fit is repeated in a moving-window fashion over the global ocean to estimate the entire mean field.

\begin{figure}
\centering
\includegraphics[width=0.76\textwidth, trim=7cm 0 0 8cm, clip = true]{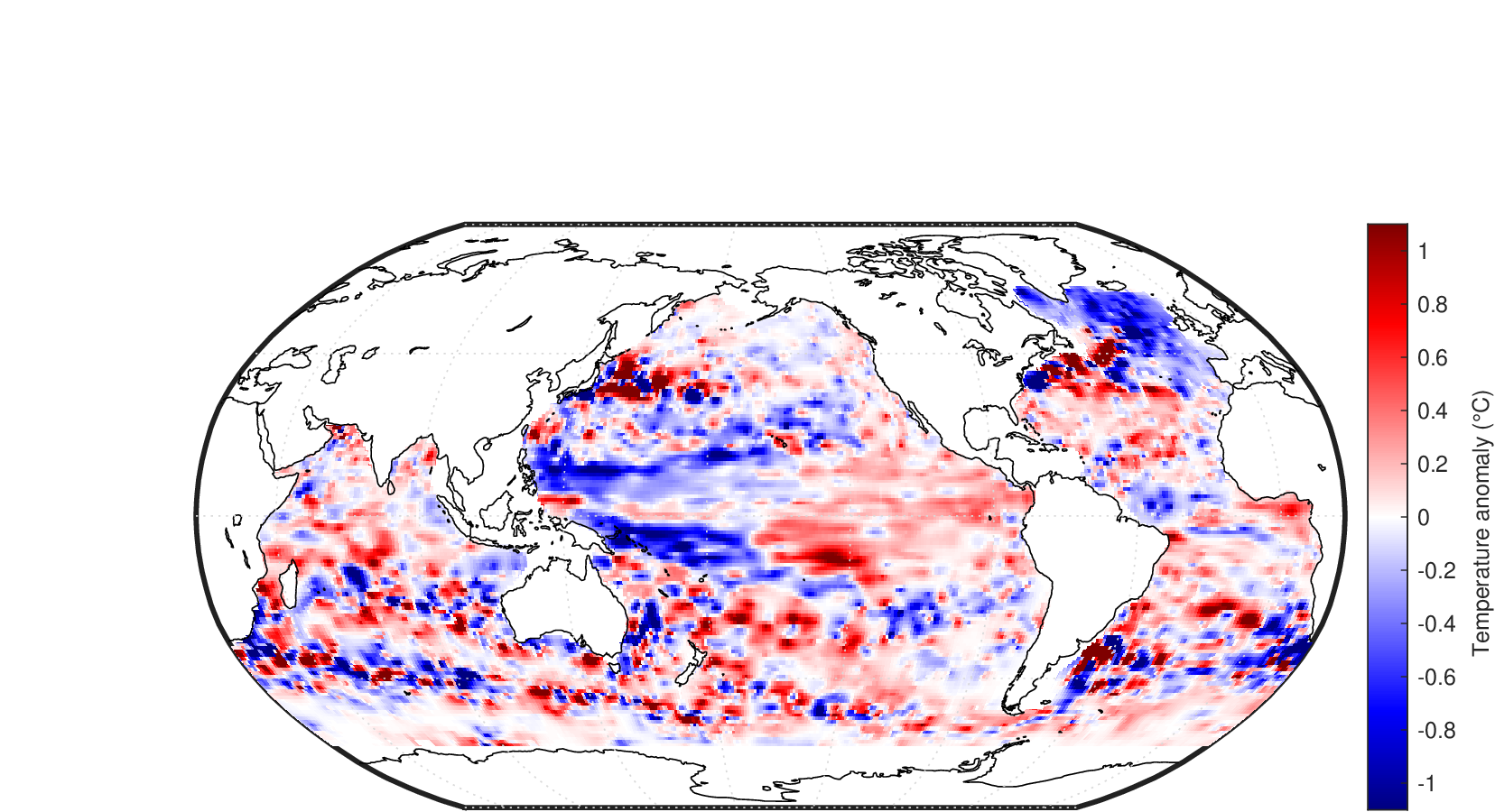}
\caption{Spatio-temporally interpolated Argo temperature anomaly field at 300 m depth in February 2015 produced using locally stationary Gaussian process regression \citep{kuusela2018locally}. The start of the 2015--16 El Ni\~{n}o event can be seen in the formation of a cold-warm dipole pattern in the Equatorial Pacific Ocean. The strong non-stationarity of the field is evident---the western sides of the ocean basins and the Southern Ocean are dominated by energetic eddies and meandering currents, while the eastern sides of the ocean basins are less dynamic and tropical regions stand out with longer-range spatial dependence.}
\label{fig:argo_anomaly}
\end{figure}

In addition to non-stationarity in the mean, the transient variation around the mean is strongly non-stationary. For example, areas with strong currents (such as western boundary currents or the Antarctic Circumpolar Current) have much more variation than less dynamic regions of the ocean. Statistically, this means that non-stationary models are needed for $a(x,y,z,t)$ and $\varepsilon(x,y,z,t)$. Again, writing down a global non-stationary model is unlikely to capture the rich structure in these fields, so a more nonparametric approach is needed. Motivated by the local regression for the mean field, \citet{kuusela2018locally} introduced a locally stationary Gaussian process (GP) regression approach for Argo mapping. In this method, one focuses on a fixed pressure level $z$ and defines a spatio-temporal moving window
$
W(x^*,y^*,t^*)=[x^*-\lambda_G,x^*+\lambda_G] \times [y^*-\lambda_G,y^*+\lambda_G] \times [t^*-\lambda_t,t^*+\lambda_t]
$
around location $(x^*,y^*,t^*)$ of size $\lambda_G$ spatially and $\lambda_t$ temporally. These authors then assume that locally within this window
\begin{equation*}
a(x,y,t) + \varepsilon(x,y,t) \overset{\text{i.i.d.}}{\sim} \mathrm{GP}(0,k((x_1,y_1,t_1),(x_2,y_2,t_2)) + \sigma^2\mathbf{1}\{(x_1,y_1,t_1)=(x_2,y_2,t_2)\}),
\end{equation*}
where $k((x_1,y_1,t_1),(x_2,y_2,t_2))$ is an anisotropic stationary covariance function and the i.i.d.~statement is over annual repetitions of the window $W(x^*,y^*,t^*)$. Treating annual anomalies as repeated observations from the same underlying Gaussian process facilitates accurate and stable estimation of the covariance parameters, which are estimated using maximum likelihood for each window $W(x^*,y^*,t^*)$ on a fine spatio-temporal grid. The estimated models are then used to obtain a kriging prediction at the center $(x^*,y^*,t^*)$ of each window. Figure~\ref{fig:argo_anomaly} shows an example of a mapped Argo temperature anomaly field produced using this approach. Since the Gaussian processes are fitted locally in space and time, this approach is computationally tractable on a global scale and is able to capture the non-stationarity of global oceanographic fields, which improves the accuracy of predictions and their uncertainties \citep{kuusela2018locally,sukianto2026locally}.

The local GP approach does not directly yield a global covariance model. An alternative that does not have this limitation is introduced in \citet{baugh2022hierarchical}, where a Gaussian process model with a global non-stationary covariance function for Argo data is created using kernel convolutions \citep{Higdon1998, paciorek2006spatial}. The key idea is that there is a kernel function associated with each spatial location $(x,y)$ that defines the local properties of the process near that location. A valid covariance function between any two locations is then given by a convolution of these kernels. Specifically, these authors consider the following covariance function for the anomaly field $a$,
\begin{align}
&k((x_1,y_1),(x_2,y_2)) \\&= \sqrt{\phi(x_1,y_1) \phi(x_2,y_2)} \int \exp(-d(x_1,y_1,u,v;\theta(x_1,y_1)) -d(x_2,y_2,u,v;\theta(x_2,y_2)))\, \mathrm{d}u \mathrm{d}v, \notag
\end{align}
where $\phi(x,y)$ is a variance field and $d(x,y,u,v;\theta(x,y))$ is a cylindrical distance metric between $(x,y)$ and $(u,v)$ that depends on the bivariate length scale field $\theta(x,y)$. Additionally, the model has a non-stationary nugget effect with variance field $\sigma^2(x,y)$. The current form of this model does not include time $t$ or depth $z$, but extensions to include them are possible. The full model is parameterized by the fields $\phi$, $\theta$ and $\sigma^2$, as well as a mean field and mean field trend. \cite{baugh2022hierarchical} take a Bayesian approach and associate Gaussian process priors for each of these fields. Posterior inference requires MCMC sampling and computational tractability is achieved using the Vecchia approximation \citep{vecchia1988estimation}.

A further approach to non-stationary covariance modeling of Argo data is explored in \citet{guinness2021gaussian} and \citet{nag2025modeling}. These authors create a non-stationary Gaussian process by warping the original domain $D$ such that the oceanographic field of interest is stationary in the warped domain $\mathcal{D}$. Let $T: \mathbb{R}^3 \rightarrow \mathbb{R}^3, (x,y,z) \mapsto T(x,y,z)$ be a sufficiently regular invertible mapping. When applied to the original domain $D$, $T$ warps the space into the new domain $\mathcal{D} = \{ T(x,y,z) : (x,y,z) \in D \}$. The key modeling assumption then is that the process $f$ has a stationary covariance in the domain $\mathcal{D}$, that is,
\begin{equation}
\mathrm{cov}(f(x_1,y_1,z_1), f(x_2,y_2,z_2)) = k(T(x_1,y_1,z_1),T(x_2,y_2,z_2)),
\end{equation}
where $k$ is a stationary covariance function. Time can also be included in the warped covariance function \citep{guinness2021gaussian}. \citet{nag2025modeling} use neural autoregressive flows as a flexible parameterization of the map $T$, while \citet{guinness2021gaussian} used gradients of spherical harmonic functions. Many other parameterizations of the warping function are available; see \citet{nag2025modeling} and the references therein for other proposed warping models. 

\subsection{Modeling the vertical dimension}

An unusual aspect of Argo data is the presence of the vertical dimension $z$. In principle, extending standard space-time models from the three-dimensional domain over $(x,y,t)$ to the four-dimensional domain over $(x,y,z,t)$ poses no issues. However, due to the physics of the ocean, the vertical dimension behaves fundamentally differently than the other dimensions. The most obvious issue is that the ocean is strongly stratified, leading to a layered structure in the vertical dimension \citep{talley2011descriptive}. The near-surface portion of the water column forms a \emph{mixed layer} where the ocean is relatively homogeneous. Below this layer, ocean properties tend to change as a function of depth, and there is typically a sharp transition between these two layers. Additionally, the depth of the mixed layer varies by season and spatial location and can vary from a few tens of meters to several hundred meters. A further complication is created by eddies which can form localized vertically correlated disturbances across a large portion of the water column.

Statistically, this means that a specialized, strongly non-stationary model is warranted for the vertical dimension. One way to proceed is to simply add the vertical dimension to a GP model over $(x,y)$ or $(x,y,t)$ and write down an appropriate non-stationary covariance model in $z$. \cite{salvana20223d} propose such a model based on a differential operator approach. The basic idea is to define the observed random field through derivatives of a collection of stationary random processes. By allowing the contribution of each derivative to vary with depth, one can create a flexible non-stationary model in the vertical dimension. The authors demonstrate that this produces plausible models of the vertical dimension in selected subregions of the Argo-sampled ocean.

An alternative approach, introduced in \citet{yarger2022functional}, is to approach the problem from the perspective of functional data analysis. This approach writes down a spatio-temporal model directly for the vertical profiles, which are understood as smooth functions of $z$. Specifically, \citet{yarger2022functional} model the anomaly field as
\begin{equation}
a(x,y,z,t) = \sum_{k=1}^K Z_k(x,y,t) \phi_k(z),
\end{equation}
where the basis functions $\phi_k(z)$ are obtained using the $K$ leading functional principal components. This reduces the problem to modeling the score fields $Z_k(x,y,t)$, which, after decorrelation, are modeled as spatio-temporal Gaussian processes with a Mat\'{e}rn covariance. The functional model is estimated locally using moving windows similar to \citet{kuusela2018locally}. One benefit of this model is that it enables inferring functionals of the mapped vertical profiles without vertical discretization and with full accounting for vertical dependence. As a concrete demonstration of this, \citet{yarger2022functional} use this approach to map the ocean heat content (vertical integral of the field) and the mixed layer depth (depth at which the vertical derivative of the density crosses a given threshold).

\subsection{Non-Gaussian modeling}

It is well-established that oceanographic fields can have non-Gaussian variation \citep[e.g.,][]{Bingham_etal2002,Sura_and_Sardeshmukh2008}. Modeling this requires considering non-Gaussian models for the anomaly field $a(x,y,z,t)$ and the nugget $\varepsilon(x,y,z,t)$ in the model of Equation~\ref{eq:basic_model}. \citet{kuusela2018locally} found that cross-validated predictive distributions of subsurface temperature exhibit heavy tails. To account for this, they proposed using a Student-$t$ distribution for the nugget effect in the moving-window Gaussian process model. The estimated nugget effect was found to be strongly heavy-tailed, especially in highly dynamic regions of the ocean, and the use of the non-Gaussian model was shown to improve the calibration of the predictive uncertainties.

Developing a non-Gaussian model for the anomaly field is statistically more challenging. \citet{bolin2020multivariate} and \citet{saduakhas2025incorporating} considered models based on stochastic partial differential equations (SPDEs) for this task. The key underlying observation is that a zero-mean random field $a$ on $\mathbb{R}^d$ with a Mat\'{e}rn covariance with length scale $\kappa$ and smoothness $\nu$ can be obtained as a solution to the SPDE \citep{whittle1963stochastic}
\begin{equation}
(\kappa^2 - \Delta)^{\alpha/2} a = \mathcal{W}, \label{eq:SPDE}
\end{equation}
where $\Delta$ is the Laplacian, $\alpha = \nu + d/2$, and $\mathcal{W}$ is a white noise process. If $\mathcal{W}$ is Gaussian, this produces a Gaussian process, but by considering non-Gaussian $\mathcal{W}$, one can define a rich class of non-Gaussian models for $a$. Specifically, \citet{bolin2020multivariate} and \citet{saduakhas2025incorporating} considered bivariate versions of this model driven by normal-inverse Gaussian white noise for Argo temperature and salinity observations. They found that the non-Gaussian SPDE models yielded improved predictions, especially in terms of metrics quantifying the shape of the predictive distributions. In these models, the nugget effect is still Gaussian. We are unaware of work on Argo data with non-Gaussian models for both the anomaly field and the nugget effect. A non-Gaussian model for $a$ could also be obtained by convolving a non-Gaussian white noise process with a kernel \citep{Higdon2002}, but the SPDE models may have computational advantages, since they enable an efficient discretization using the finite-element method \citep{lindgren2011explicit, bolin2020multivariate}.

\subsection{Multivariate modeling} \label{sec:multivariate}

The various oceanographic fields do not evolve in isolation, but tend to be dependent due to the fundamental physical laws that govern ocean dynamics. Statistically, this means that the various fields have non-zero cross-correlations, which presents an opportunity to use multivariate spatio-temporal modeling to borrow strength across the fields to obtain improved interpolations. This is particularly true when some fields are observed more densely than others, in which case the densely observed fields can be leveraged to improve the mapping of the more sparsely observed ones.

Most of the previously discussed mapping methods have been extended to the multivariate setting. For simplicity, we focus here on the bivariate case; generalizations to more than two fields are relatively straightforward. The bivariate version of Equation~\ref{eq:basic_model} is
\begin{equation}
\begin{bmatrix} f_1(x,y,z,t) \\ f_2(x,y,z,t) \end{bmatrix} = \begin{bmatrix} \mu_1(x,y,z,t) \\ \mu_2(x,y,z,t) \end{bmatrix} + \begin{bmatrix} a_1(x,y,z,t) \\ a_2(x,y,z,t) \end{bmatrix} + \begin{bmatrix} \varepsilon_1(x,y,z,t) \\ \varepsilon_2(x,y,z,t) \end{bmatrix},
\end{equation}
where $\mu_i$, $a_i$ and $\varepsilon_i$ are the mean, anomaly and nugget fields of the $i$th oceanographic field~$f_i$. Moving from the univariate model to the bivariate version does not necessarily create any fundamental difficulties in estimating the marginal mean fields $\mu_1$ and $\mu_2$. Both can still be estimated separately using local regression or other similar techniques, although one may wish to use the more densely observed mean field as a covariate in the regression model for the more sparsely observed field. The bigger difference from the univariate model is that now the anomaly fields $a_1$ and $a_2$ can be cross-correlated, which can be characterized using the spatio-temporal cross-covariance function
\begin{equation}
\mathrm{cov}(a_1(x_1,y_1,z_1,t_1), a_2(x_2,y_2,z_2,t_2)) = k_{12}((x_1,y_1,z_1,t_1),(x_2,y_2,z_2,t_2)).
\end{equation}
The nugget effects are still uncorrelated across $x$, $y$ and $t$, but might be correlated between the fields at the same spatio-temporal location, so for models without the vertical dimension,
\begin{equation}
\mathrm{cov}(\varepsilon_i(x_1,y_1,t_1), \varepsilon_j(x_2,y_2,t_2)) = \begin{cases} \sigma_i^2, &\text{ if } (x_1,y_1,t_1) = (x_2,y_2,t_2) \text{ and } i=j, \\ \rho \sigma_1 \sigma_2, &\text{ if } (x_1,y_1,t_1) = (x_2,y_2,t_2) \text{ and } i \neq j, \\ 0, &\text{ if } (x_1,y_1,t_1) \neq (x_2,y_2,t_2), \end{cases}
\end{equation}
where $\sigma_1^2$ and $\sigma_2^2$ are the marginal nugget variances and $\rho$ is the nugget cross-correlation. A non-zero nugget cross-correlation implies that the two fields are correlated at spatio-temporal scales smaller than the Argo sampling resolution. This is likely true for most oceanographic fields due to eddies and other similar mesoscale and submesoscale processes. \cite{saduakhas2025incorporating} study the importance of having a cross-correlated nugget effect in a bivariate model for Argo temperature and salinity and conclude that doing so leads to substantially better model fits and spatial predictions. When the model includes the vertical dimension, it can be appropriate to consider a vertically correlated bivariate nugget effect, as in \citet{yarger2022functional} and \citet{sukianto2026improved}.

The different multivariate mapping approaches for Argo data primarily differ in terms of the modeling of the cross-covariance $k_{12}$ for the anomaly fields. Since it is common to model the marginal fields using Mat\'{e}rn covariances with different anisotropic length scales, cross-covariance models that are compatible with this marginal specification are particularly appealing. \citet{gneiting2010matern} provide a characterization of a broad class of such models. A particularly interesting variant is the cross-covariance function defined in \citet{Kleiber2012}. \citet{sukianto2026improved} use a moving-window version of that model in a bivariate extension of the locally stationary GP approach of \citet{kuusela2018locally} to map Argo ocean heat content jointly for two vertical sections of the ocean. They use a paired cross-validation approach to show that this produces accurate predictive cross-correlations across the two fields and demonstrate that the bivariate approach improves the maps and the quantification of their uncertainties.

A further way to obtain a bivariate model with Mat\'{e}rn margins is to consider a bivariate extension of the SPDE model in Equation~\ref{eq:SPDE}. \citet{bolin2020multivariate} and \citet{saduakhas2025incorporating} define a bivariate process as the solution to the system of SPDEs
\begin{equation}
\bm{D}
\begin{bmatrix}
(\kappa_1^2 - \Delta)^{\alpha_1/2} & 0 \\
0 & (\kappa_2^2 - \Delta)^{\alpha_2/2}
\end{bmatrix}
\begin{bmatrix}
    a_1 \\ a_2
\end{bmatrix}
 =
\begin{bmatrix}
\mathcal{W}_1 \\ \mathcal{W}_2    
\end{bmatrix},
\end{equation}
where $\bm{D}$ is an invertible $2 \times 2$ matrix, $\alpha_i = \nu_i + d/2$, $\kappa_i$ and $\nu_i$ are the length scale and smoothness of marginal field $i$ and $\mathcal{W}_i$ are uncorrelated white noise processes. Here, the choice of $\bm{D}$ controls the dependence between the two fields. When $\mathcal{W}_i$ are Gaussian, this produces a bivariate GP with Mat\'{e}rn margins, but one also has the flexibility of defining a non-Gaussian bivariate process by driving the SPDE with non-Gaussian noise. \cite{bolin2020multivariate} find that a bivariate non-Gaussian SPDE model with an uncorrelated nugget improves the interpolation of Argo temperature fields, while \cite{saduakhas2025incorporating} conclude that in the presence of a correlated nugget, the benefits of the non-Gaussian model are less clear for modeling Argo temperature and salinity. Further multivariate model constructions proposed for Argo data include the differential operator approach in \citet{salvana20223d}, functional data approaches in \citet{yarger2022functional} and \citet{korte2025functional}, confluent hypergeometric covariances in \citet{yarger2025multivariate} and asymmetric multivariate Mat\'{e}rn covariances in \citet{yarger2026multivariate}.

A particularly compelling use of multivariate modeling in the context of Argo data is in mapping sparsely observed BGC-Argo variables with the help of the more densely observed Core Argo variables. \citet{korte2025functional} develop a spatio-temporal functional data analysis approach to jointly model Argo temperature, salinity and oxygen profiles in the Southern Ocean. Their approach enables the production of oxygen maps that combine information from the more densely observed temperature and salinity profiles with the sparser oxygen profiles. Their approach also uses probabilistic clustering to divide the Southern Ocean into distinct dynamic regions to accommodate non-stationarity caused by ocean fronts. An alternative way to map oxygen based on temperature and salinity using machine learning regression models is studied in \citet{giglio2018estimating} and \citet{sharp2023gobai}.

\section{Statistical analysis of surface drifter data}\label{ss:drifters}

 As with profiling floats, drifter data could be treated as spatio-temporal Eulerian data, where the drifters sample continuously evolving spatial phenomena (such as velocity or temperature fields) at locations that change in time as the drifters move. In this Eulerian perspective, we can then perform local spatial predictions in space-time near drifter locations using statistical techniques such as kriging or Gaussian process (GP) regression \citep{xia2025error}, similar to those discussed in Section~\ref{ss:floats} for profiling floats, or we can fit drifter data to local Eulerian decompositions of the flow field from fluid dynamics models \citep{ponte2024inferring}.

In practice, however, large-scale spatial predictions are unfeasible owing to the sparsity of drifter deployments with respect to the sheer size of the ocean (in the Global Drifter Program, there are only 1,000--1,500 drifters deployed at once in practice, see Figure~\ref{fig:locations}), with huge gaps yielding large prediction errors unless covariate information (e.g., from satellites) is integrated (see Section~\ref{ss:fusion} for discussion of data fusion).

Instead, the richness in drifter data lies in the temporal frequency of each drifter's measurements (hourly for the Global Drifter Program), rather than their spatial coverage. 
For this reason, drifter data are often modeled as time series of particle trajectories moving in space over time, rather than as sparse spatial point measurements of some large (random) field. This, in essence, suggests a Lagrangian treatment of such data, where the frame of reference is the drifter, not some fixed location, as discussed in Section~\ref{ss:frame}. This allows drifter trajectories to be modeled as stochastic processes in time, as we shall now discuss.

\subsection{Stochastic process representations of drifter motion}\label{ss:stoch}

The stochastic modeling of Lagrangian particles goes back to \cite{taylor1922diffusion} and \cite{kolmogorov1941local}. An excellent starting point to synthesizing these approaches is \cite{berloff2002material}, who propose a hierarchy of stochastic differential equation representations. Here we will follow \cite{lacasce2008statistics} and describe this hierarchy in a stationary and homogeneous setting for simplicity (where parameters do not change in space and time), but the interested reader can refer to \cite{berloff2002material} for more general expressions. The simplest base model is a `Markov-0' model, which models the random displacement of a drifter (or some Lagrangian particle in general) in the $x$-direction via
\begin{equation}
    \mathrm{d}x = \overline{u}\mathrm{d}t + \sqrt{2\sigma^2} \mathrm{d}W,
    \label{eq:Markov0}
\end{equation}
and similarly in the $y$-direction, where $\overline{u}$ is the spatially homogeneous mean velocity (called the mean flow), $W$ is a Wiener process (or Brownian motion) satisfying $W(t+\Delta)-W(t)\sim\mathcal{N}(0,\Delta)$ for $\Delta>0$, and $\sigma^2>0$ is the {\em diffusivity}, defined by \cite{taylor1922diffusion} as one half of the time derivative of the {\em dispersion} (i.e., the displacement variance), where for the Markov-0 model the dispersion is $\mathrm{Var}\{x(t)-x(0)\}=2\sigma^2t$ for $t>0$. The Markov-0 model 
is a continuous-time random walk with drift in stochastic processes, and is the Lagrangian (particle) representation of the classical advection–diffusion equation in fluid dynamics, where the mean flow governs the (deterministic) advection and the scaled increments of the Wiener process create the (random) diffusion. Further discussion can be found in \cite{lacasce2008statistics} and \cite{van2018lagrangian}, with the Markov-0 model considered a useful, but ultimately poor, baseline model for fluid particle motion due to two limitations: (1) velocities are `delta-correlated' (i.e., a white noise process), and (2) the dispersion grows linearly in time at all timescales---both considered unrealistic for drifter motion.

A more realistic and widely-used model is the `Markov-1' (or {\em first-order} Markov) model, which simultaneously allows for velocity autocorrelation and non-linear growth in dispersion by going to the next order of differentiation, i.e.,
\begin{equation}
    \mathrm{d}x = (u+\overline{u})\mathrm{d}t,\quad \mathrm{d}u=-\frac{1}{T_L}u\mathrm{d}t+\sqrt{\frac{2\sigma^2}{T_L}} \mathrm{d}W,
    \label{eq:Markov1}
\end{equation}
and similarly in the $y$-direction, where $T_L>0$ is known as the Lagrangian (or integral) timescale parameter, $\sigma^2=\mathrm{Var}\{u(t)\}>0$ sets the variance of the stationary velocity process, and $\overline{u}$ and $W$ are as in the Markov-0 model of Equation~\ref{eq:Markov0}. The interpretation, and hence the units, of $\sigma^2$ differ between the Markov-0 and Markov-1 models (diffusivity has units of $m^2/s$, velocity variance has units of $m^2/s^2$). The interpretation of $T_L$ becomes clearer through the autocorrelation function of the velocity process, which for a Markov-1 model is $\rho(\tau) = \exp(-|\tau|/T_L)$, such that $T_L=\int_0^\infty\rho(\tau)\mathrm{d}\tau$ governs the rate of autocorrelation decay. With a Markov-1 model, the dispersion grows quadratically with time at short timescales (a {\em ballistic} regime) before transitioning to linear growth over long timescales (a {\em diffusive} regime), consistent with fundamental theory on diffusion \citep{taylor1922diffusion}. The long-term dispersion growth is $\mathrm{Var}(x(t)-x(0))\sim2\sigma^2T_Lt$, such that the long-term diffusivity is $\sigma^2T_L$. The Markov-1 model corresponds to an Ornstein-Uhlenbeck velocity process with constant mean in stochastic processes, and a Langevin model in fluid dynamics. It has been routinely compared with drifter data as a benchmark model, for example in \cite{koszalka2013pursuit}.

The `Markov-2' (or {\em second-order} Markov) model goes to the next order and allows for autocorrelation in the particle accelerations $a(t)$ via the model
\begin{equation}
    \mathrm{d}x = (u+\overline{u})\mathrm{d}t,\quad \mathrm{d}u=-\frac{1}{T_v}u\mathrm{d}t+a\mathrm{d}t, \quad \mathrm{d}a = -\frac{1}{T_a}a\mathrm{d}t+\left(\frac{\sqrt{2\sigma^2(T_v+T_a)}}{T_vT_a}\right) \mathrm{d}W, 
\end{equation}
and similarly in the $y$-direction, where $T_v>0$ and $T_a>0$ are the decay timescales of the velocity and acceleration processes respectively, and $\sigma^2>0$ defines the stationary velocity variance as with the Markov-1 model. The Markov-2 model captures more complex geophysical fluid structure than the Markov-0 and Markov-1 models, but a practical question is whether such structure can realistically be resolved from drifter data. A study by \cite{rupolo2007lagrangian} on Global Drifter Program drifters found that $T_a$ is better resolved along major currents such as the Gulf Stream, but less so in regions away from major currents. \cite{berloff2002material} also propose a Markov-3 model, of which we omit the details, and in practice would be difficult to resolve with drifter data.

Several extensions to this hierarchical framework have been proposed to more realistically mimic drifter motion. In particular, drifters tend to exhibit swirling and oscillatory motion, due to eddies, tides and Coriolis-effect induced inertial oscillations \citep{early2012forces}. Motivated by this, instead of requiring higher-order Markov models to induce oscillations, a simple extension of the Markov-1 model of Equation~\ref{eq:Markov1} was proposed by \cite{veneziani2004oceanic} for vortices and eddies of sub-surface drifters given by
\begin{equation} \label{eq:veneziani}
    \mathrm{d}u=-\frac{1}{T_X}u\mathrm{d}t+\Omega v \mathrm{d}t+\sqrt{\frac{2\sigma_X^2}{T_X}} \mathrm{d}W_X, \quad \mathrm{d}v=-\frac{1}{T_Y}v\mathrm{d}t-\Omega u \mathrm{d}t+\sqrt{\frac{2\sigma_Y^2}{T_Y}} \mathrm{d}W_Y,
\end{equation}
where $T_X,T_Y>0$, $\Omega\in\mathbb{R}$ and $\sigma^2_X,\sigma^2_Y > 0$, such that the zonal (eastward) and meridional (northward) velocities, $u$ and $v$, are no longer independently governed, and instead exhibit rotational motion in the Cartesian plane with frequency $|\Omega|/(2\pi)$. This model was also proposed for inertial oscillations in surface drifters by \cite{sykulski2016lagrangian} where $\Omega = (2 /T)\sin\psi$, with $\psi$ the latitude of the drifter and $T$ one sidereal day (23h 56min 4.1s). Therefore at latitude 30 degrees the inertial oscillation has a period of exactly one sidereal day, but will be faster towards the poles and slower towards the Equator. In addition to the model of Equation~\ref{eq:veneziani} for inertial oscillations, \cite{sykulski2016lagrangian} also proposed adding an independent Mat\'ern process (see also \cite{lilly2017fractional,guillaumin2017analysis}) to capture background turbulent motion (see Section~\ref{ss:spectra} for more discussion).

A number of procedures have been proposed to estimate parameters of such models from drifter data. \cite{rupolo2007lagrangian} estimates parameters directly from the sample velocity and acceleration variances, and by using an estimate of the Lagrangian timescale ($T_L=\int_0^\infty\rho(\tau)\mathrm{d}\tau$) by empirically estimating the velocity autocorrelation function $\rho(\tau)$ and then truncating the integral by integrating from zero until the first `zero-crossing' to avoid noise issues at high lags (a widely used, but perhaps naive, estimator across oceanography \citep{lumpkin2002lagrangian}). \cite{veneziani2004oceanic} adopts a similar approach of equating the empirical velocity autocorrelation to the parametric form of Equation~\ref{eq:veneziani} by carefully selecting certain lags of interest (in essence a moment-matching procedure). \cite{sykulski2016lagrangian} and \cite{sykulski2019debiased} propose maximum likelihood procedures in the frequency domain using Fast Fourier Transforms (known as `Whittle' methods), and perform these fits semi-parametrically to avoid frequencies in the spectrum where the model is misspecified.

In recent years there have also been several developments in using modern machine learning approaches to predict or interpolate gappy drifter trajectory data using neural networks \citep{brolly2023inferring,grossi2025surface}, random forest and support vector regression \citep{medina2025using}, or generative diffusion models \citep{li2025stochastic} for example. Comparisons were made with ARIMA and Gaussian process regression in these papers, but this is an area of very active research to compare, and potentially combine, a broader class of data-driven and physics-informed stochastic models.

\subsection{Spectra and wavelets} \label{ss:spectra}
Spectra are of broad interest in fluid dynamics to characterize different scaling behaviors in frequency (time) and wavenumber (space). With Lagrangian data we can compute the {\em Lagrangian frequency spectrum}, which for a stationary and homogeneous flow field is equal to the Fourier transform of the velocity autocorrelation function $\rho(t)$, i.e.,
\begin{equation}
    f(\omega) = \sigma^2 \int_{-\infty}^\infty \rho(\tau) \exp(-\mathrm{i}2\pi\omega\tau)\mathrm{d}\tau = 2\sigma^2 \int_0^\infty \rho(\tau) \cos(2\pi\omega\tau)\mathrm{d}\tau,
\end{equation}
where $\mathrm{i}^2=-1$ and $\sigma^2$ is the velocity variance. The Lagrangian frequency spectrum, $f(\omega)$, can therefore be interpreted as the distribution of velocity variance over frequency. The Markov hierarchy introduced in Section~\ref{ss:stoch} has a natural link to spectra, with $f(\omega)$ constant for a Markov-0 model (as the velocities behave as white noise), but for large $\omega$, $f(\omega)\sim \omega^{-2}$ for a Markov-1 model, $f(\omega)\sim \omega^{-4}$ for a Markov-2 model, and so on \citep{berloff2002material}, such that the spectra decay according to a power law. The Markov hierarchy therefore only characterizes even-integer decay rates in the Lagrangian frequency spectrum. However, in many observational and theoretical studies the decay rate of the Lagrangian frequency spectrum sits between the even integers (e.g., $f(\omega)\sim \omega^{-3.3}$ is observed in \cite{rupolo1996lagrangian}). This motivates the use of the Mat\'ern process for Lagrangian velocities, as this creates a power spectrum of the form $f(\omega)\sim \omega^{-\alpha}$ ($\alpha>1$) as $\omega$ grows, allowing a full continuum between the Markov models \citep{lilly2017fractional}. The parameter $\alpha$ controls the smoothness of the velocity process, with larger values of $\alpha$ corresponding to smoother velocity processes and, after integration, smoother displacement trajectories. The Mat\'ern process can be extended to anisotropic versions with preference for zonal or meridional random dispersion at different frequencies \citep{sykulski2017frequency}. The use of the Mat\'ern to model atmospheric and oceanic turbulence has a rich history \citep{guttorp2006studies}. 

Stochastic processes that exhibit oscillations, such as the Markov-1 extension of Equation~\ref{eq:veneziani} \citep{veneziani2004oceanic}, will exhibit a peak in the spectrum at frequency $|\Omega|/(2\pi)$. As drifters record velocities in two spatial dimensions (zonal and meridional) then rather than modeling these separately, oceanographers often combine them to form two rotary spectra \citep{poulain2013mediterranean,yu2019surface,arbic2022near} which decompose velocity variance in terms of clockwise and anti-clockwise rotations in two-dimensional space. For example, the \cite{veneziani2004oceanic} model will exhibit a peak at frequency $|\Omega|/(2\pi)$ in one of these spectra and not the other, depending on whether $\Omega$ is positive or negative (determined by whether the drifter is in the northern or southern hemisphere). A global analysis of rotary spectra on Global Drifter Program data can be found in \cite{elipot2008spectral}.

When studying drifter velocities, it is recognized (see, for example, \cite{lacasce2008statistics}), that there should be a separation 
timescale for the mean flow variation and for any residual variation. One might in this setting decompose the observed velocities $\{u(t),v(t)\}$ into unobserved components representing the mean and residuals respectively,
\begin{equation}
\begin{pmatrix}
u(t)\\
v(t)
  \end{pmatrix}=\begin{pmatrix}
u^{(m)}(t)\\
v^{(m)}(t)
  \end{pmatrix}+\begin{pmatrix}
u^{(o)}(t)\\
v^{(o)}(t)
  \end{pmatrix}.
  \label{eqn:unobscomp}
\end{equation}
However, because drifters follow currents, they move across different parts of the ocean, so all components of the velocity process are likely to be non-stationary. The mean component $\{u^{(m)}(t),v^{(m)}(t)\}$ can be estimated using spatial and temporal binning of drifters, using techniques such as Gauss--Markov estimation to account for autocorrelation, see e.g. \cite{lumpkin2013global,qian2014estimation} and \cite{laurindo2017improved} who also split the velocities into further components including seasonal ones. The residual component $\{u^{(o)}(t),v^{(o)}(t)\}$ will contain diffusive and oscillatory behavior at different scales, caused by eddies, tides, and inertial oscillations, among other effects.

\begin{figure}
\centering
\begin{minipage}[c]{0.45\textwidth}
    \centering
    \includegraphics[width=\linewidth]{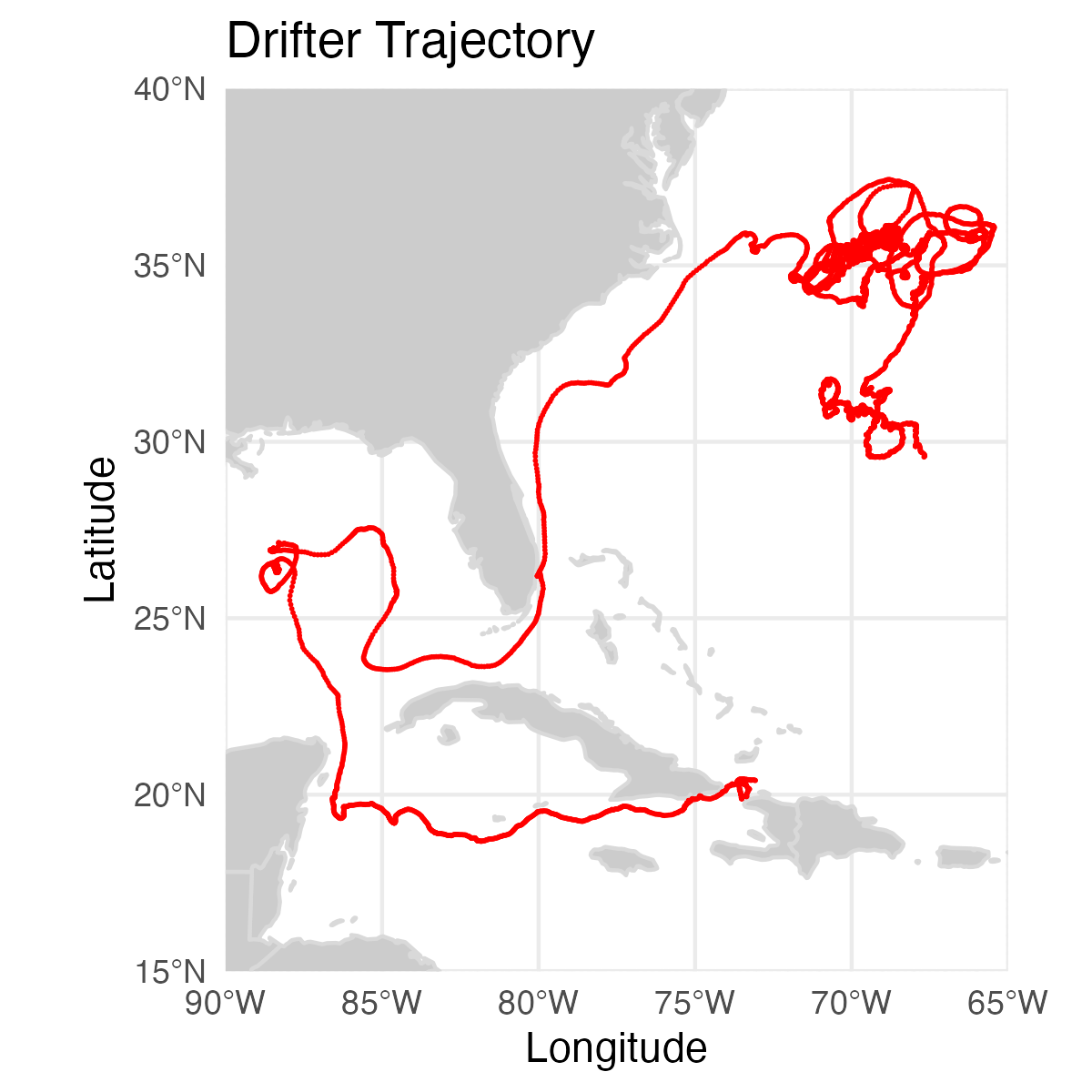}
\end{minipage}
\hfill
\begin{minipage}[c]{0.44\textwidth}
    \centering
    \includegraphics[width=\linewidth]{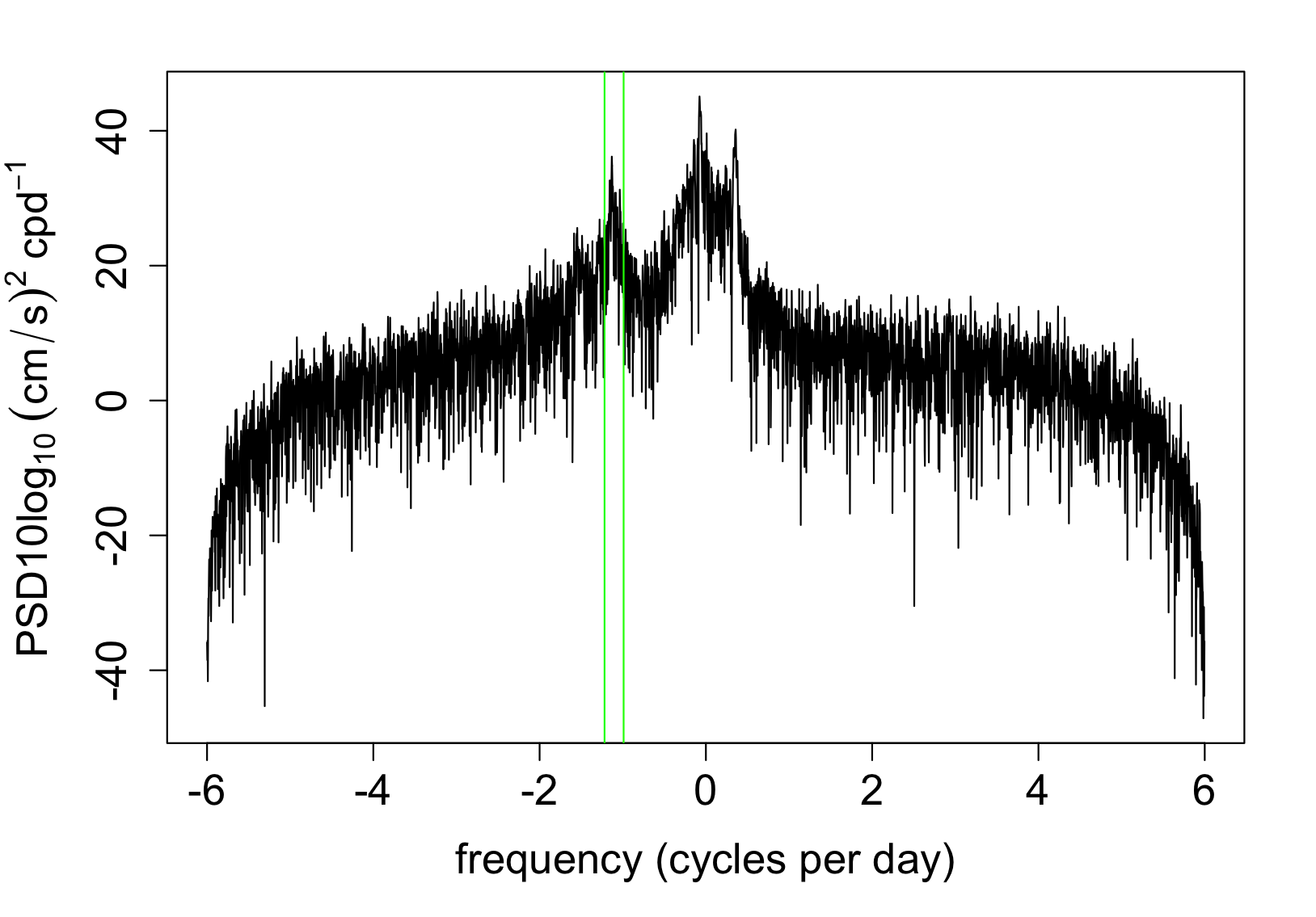}

    \vspace{-1cm}

    \includegraphics[width=\linewidth]{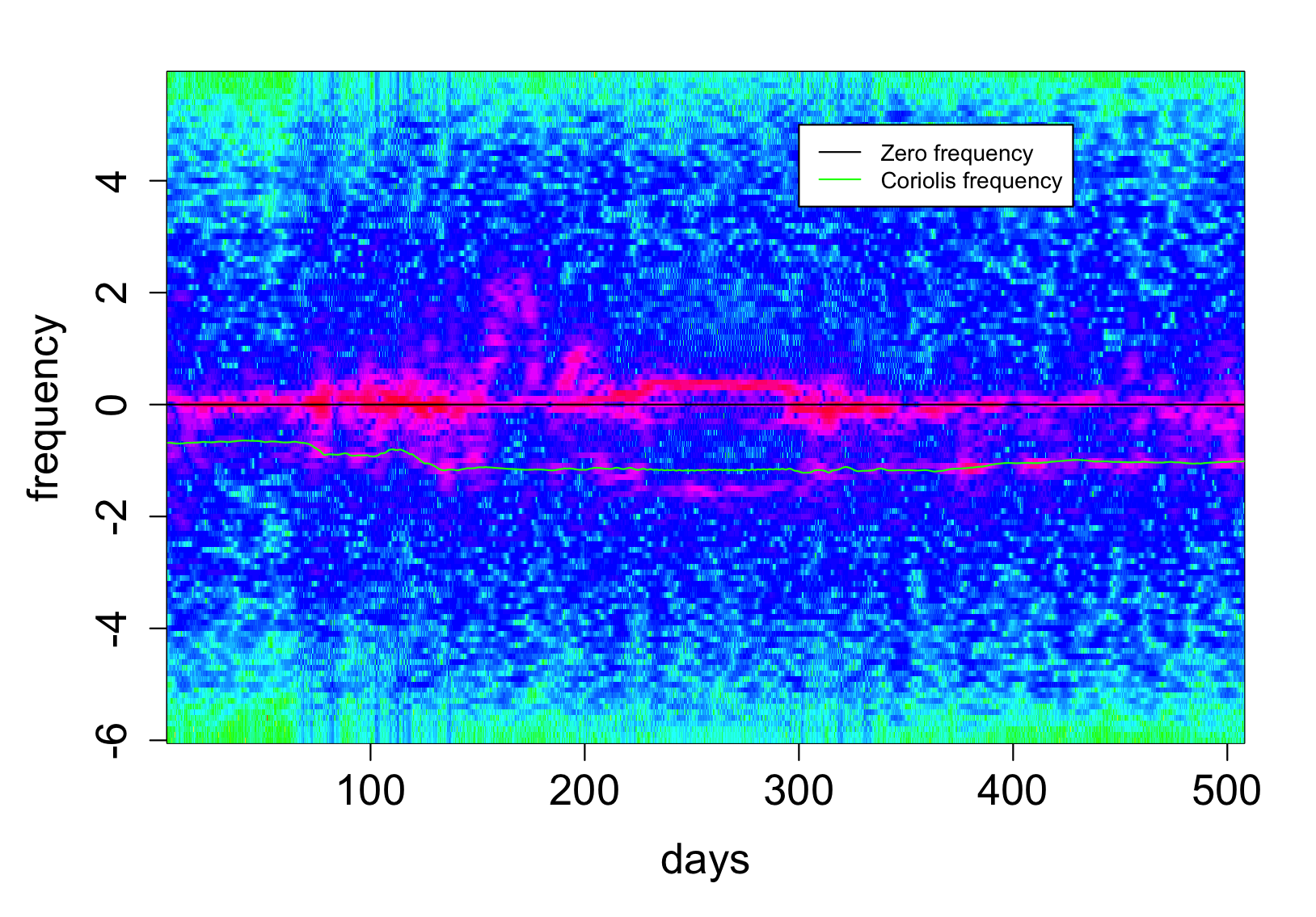}
\end{minipage}
\caption{The trajectory of drifter ID \#44000 lasting 513 days and plotted in red (left panel). A tapered power spectral density estimate of the Lagrangian frequency spectrum of the drifter from day 200 onwards (upper-right panel). The taper used is a single discrete prolate spheroidal sequence taper of bandwidth 1. The green lines correspond to the minimum and maximum Coriolis frequency experienced by the drifter in this time interval. A spectrogram (or short-term Fourier transform) of the Lagrangian frequency spectrum of the drifter (lower-right panel), calculated using rolling windows of length 5 days.}
\label{fig:betty}
\end{figure}

We now illustrate such behavior on a single drifter in Figure~\ref{fig:betty}. The non-stationarity of the drifter velocities is already apparent from the spatial trajectory (left panel). The Lagrangian frequency spectrum (upper-right panel) is shown as rotary spectra; negative frequencies correspond to clockwise rotations, and positive frequencies to anti-clockwise rotations. The `spectrogram' (or short-term Fourier transform, lower-right panel) captures the non-stationarity of the drifter velocities via time-variation of the Lagrangian frequency spectrum, using rolling windows of length 5 days. It is clear from this figure that different physical processes operate over different ranges of frequencies. Changes in the overall meandering of the drifter correspond to low-frequency, red-noise-like behaviour. There are also oscillations leading to bumps in the spectrum. We can recognize more than one distinct bump---one corresponding to the clockwise inertial oscillations (located between the two green vertical lines in the upper-right panel), and the second due to the drifter being caught in an anti-clockwise eddy between days 200 and 300, which causes the bump at positive frequencies at around 0.2 cycles per day. 

One way to separate these scales from observations is to filter trajectories by distinct frequencies of the velocity process. This method has been adopted by~\cite{ivanov2019estimation,lilly2011extracting} and \cite{lilly2011analysis} using wavelets, and  by \cite{sykulski2016lagrangian,grainger2023multivariate} using short-term Fourier transforms, as in Figure~\ref{fig:betty} (lower-right panel). \cite{lilly2011extracting} and \cite{lilly2011analysis} extract oscillations where, at each point in time, an estimate of the oscillation is obtained using the wavelet ridge method~\citep{lilly2006wavelet}, namely extracting the estimate in terms of its amplitude and phase derived from the complex wavelet transform. The remaining variation can be extracted by subtracting the estimated oscillation from the aggregated signal. Properties of particles trapped in eddies can then be directly aggregated~\citep{lilly2011extracting}. 

Of interest to oceanographers are also {\em Eulerian energy spectra} which characterize the distribution of energy at fixed locations over different wavenumbers. Computing such spectra from drifters is highly nontrivial due to the way drifters move in space and time, but can be done using pairs or clusters of approximately co-located drifters; see \cite{lacasce2016estimating} and \cite{poje2017evidence} for details. These studies are also typically focussed on estimating the {\em second-order structure} function \citep{beron2016statistics} which is equivalent to the variogram used in spatial statistics, and is hence related to the Eulerian energy spectra via a Fourier transformation of the spectrum to a covariance function and then mapping the covariance function to a second-order structure function (or variogram). Also to be estimated is the finite-scale Lyapunov exponent (FSLE), which quantifies how rapidly two drifters separate by measuring the time required for their distance to increase from one finite scale to a larger finite scale \citep{lumpkin2010surface}. These quantities are often compared with fundamental geophysical fluid laws of turbulence.

\subsection{Drifter aggregations and pathway finding} 
Sections~\ref{ss:stoch} and \ref{ss:spectra} focused on understanding the stochastic nature of individual trajectories, and relating them to fundamental geophysical fluid dynamics principles. In this section we focus on data-driven approaches to understanding {\em collections} of drifter trajectories, such as from the Global Drifter Program (as in Figure~\ref{fig:spaghetti}), to understand how the ocean is connected. A fundamental approach was developed by \cite{maximenko2012pathways} who proposed discretizing the ocean into spatial bins and then estimating the {\em transition probability} of a drifter's location between pairs of spatial bins over some time horizon $T$. A typical choice of $T$ is 5 days, to exceed various oscillatory timescales and capture long-term drift behavior. The choice of spatial binning resolution is the classic bias-variance tradeoff: small bins can lead to data sparsity (high variance) and large bins can smooth over various meandering currents and pathways (high bias)---\cite{maximenko2012pathways} adopt $1/2\times1/2$ degree bins in latitude-longitude, \cite{rypina2017multi} $1 \times1$ degree bins.

The transition probabilities can be used for a multitude of purposes. Typically a Markov approximation is imposed at lag $T$, so the conditional probability of a drifter being in a spatial bin at time $t$ given that it was in a spatial bin at time $t-T$ is independent of where it was before time $t-T$. From this assumption we can use the theory of Markov chains to construct transition matrices which can be repeatedly applied to find drifter pathways, their travel times, and their spatial distributions over time. For example, Dijkstra's algorithm can be used to find the most likely pathway taken by a drifter from one location to another \citep{o2021estimating}.

\cite{maximenko2012pathways} applied the transition probability idea to track marine debris such as plastic (see also \cite{van2012origin} and \cite{miron2021transition}) where drifters are treated as a proxy for marine debris, and this idea has been applied to other ocean-borne objects such as plankton \citep{laso2023holistic}, oil spills \citep{olascoaga2023exploring} and the missing MH-370 airliner \citep{miron2019markov}. An innovation in \cite{o2021estimating} was to replace latitude-longitude spatial binning with hexagonal binning using Uber's H3 index (in essence a scaled-up version of the stitching found on certain footballs), which creates bins of approximately equal size, and was found to yield more stable transition probability estimates and pathways with more realistic `meandering' geometry. In contrast, latitude-longitude binning was found to skew probability transitions in favor of north-south or east-west transitions (like a rook in the game of chess), but penalize diagonal moves (like a bishop), thus creating blocky straight-line pathways. An online tool for finding such pathways on a hexagonal grid can be found on the Global Drifter Program website at \url{https://www.aoml.noaa.gov/phod/gdp/derived_pathways.php}.

A complementary approach is to simulate vast numbers of Lagrangian trajectories from velocity fields (many more than there are actual drifter trajectories), and then aggregate outputs to calculate statistics of interest such as travel times and streamfunctions (see \cite{van2018lagrangian} for a comprehensive review). A popular Python package for simulating such trajectories is called `Parcels' \citep{delandmeter2019parcels}. Many other statistical approaches to aggregating drifter data have been proposed, including spatial clustering of drifters to compute diffusivity \citep{koszalka2011surface, poje2014submesoscale,oscroft2020separating}, data assimilation of drifter data to improve Eulerian estimates of the flow field \citep{molcard2003assimilation}, and binning and averaging approaches to produce surface current maps \citep{poulain2013mediterranean,lilly2021gridded}.

\section{Statistical methods for data integration}\label{sec:dataintegration}
We have described statistical oceanography for profiling floats (Section~\ref{ss:floats}) and surface drifters (Section~\ref{ss:drifters}). Now we look at how data from these instruments has been integrated with that from other sources, especially satellites, as part of the GOOS. We focus on instrument calibration and validation (Section~\ref{ss:calibration}), data fusion (Section~\ref{ss:fusion}), and data assimilation (Section~\ref{ss:assimilation}).
\subsection{Instrument Calibration and Validation}\label{ss:calibration}
Surface drifters have been used to calibrate velocities derived from high-frequency radar measurements and satellite altimetry, as well as satellite-derived SST (see \cite{lumpkin2017advances} for a review). A complication is that satellite-derived velocities come from SSH measurements, which are converted to velocities by taking gradients (and adjusting for the Coriolis force), to obtain {\em geostrophic} velocities. These are known to differ from surface velocities from drifters because the latter will also experience effects from forces such as surface winds (known as Ekman effects), tides, and inertial oscillations. To account for this, one common approach is to low-pass filter drifter velocities to make them more like geostrophic velocities \citep{mulet2021synergy}, another is to build an additive model to decompose drifter velocities into various components including geostrophic velocity \citep{ballarotta2022improved}. Drifters are also used to calibrate and validate high-resolution global models for surface velocities including GlobCurrent \citep{rio2014beyond}, which also integrates Argo data.

Near-surface Argo observations can also be used to calibrate and validate satellite observations. Argo data have been particularly useful in validating remotely sensed sea surface salinity observations \citep[e.g.,][]{abe2014evaluation}. One challenge arising in this context is that most satellite observations are area averages over the ``footprint'' of the satellite instrument while in situ data are point observations. In situ data therefore contain subfootprint variability missing from the satellite observations. Statistically, this is an example of a \emph{change-of-support} problem \citep{gelfand2010misaligned}, while the geophysical literature typically uses the term \emph{representation error}. A non-negligible representation error implies an inherent limitation in using in situ data to calibrate and validate remotely sensed observations. \cite{bingham2021matchup} use simulated ocean data to analyze the magnitude of the representation error when comparing Argo salinity data with satellite sea surface salinity observations and show that it depends on how the satellite data is matched with the in situ data.

\subsection{Data Fusion}\label{ss:fusion}
Profiling floats and surface drifters are key components of the GOOS; but satellite data often provide better spatial and temporal coverage. It is natural to integrate and fuse these data sources, and there has been considerable recent progress in doing so in a data-driven fashion using machine learning. For example, drifter data has been fused with satellite-derived estimates of SSH, SST, chlorophyll and surface winds. Typically the goal is to predict or forecast currents in a region \citep{garcia2025orcast,kugusheva2024ocean,o2021estimating}, or the path of a drifter trajectory \citep{aksamit2020machine,tian2025prediction}. The machine learning techniques used include recurrent neural networks \citep{aksamit2020machine},
convolutional neural networks \citep{kugusheva2024ocean,tian2025prediction} and natural gradient boosting \citep{o2023probabilistic}.

Similarly, satellite observations have been fused with Argo data to obtain higher-resolution maps of subsurface variables. For example, \cite{lyman2023global} train a random forest regression model to predict subsurface ocean heat content anomalies (OHCA) based on high-resolution satellite SSH and SST observations. The regression model used is
\begin{equation}
    \widehat{\text{OHCA}} = f(x,y,t,\text{SSH},\text{SST}), \label{eq:rf_regression}
\end{equation}
where $\widehat{\text{OHCA}}$ is the predicted anomaly at longitude $x$, latitude $y$ and time $t$ and $f$ is a random forest trained using Argo OHCA observations with colocated satellite SSH and SST. Once the model is trained, it can be applied to high-resolution SSH and SST maps to obtain a high-resolution OHCA map. \cite{lyman2023global} show that the resulting maps provide improved consistency with top-of-atmosphere energy flux time series and are able to resolve eddy-scale features, which is impossible using in situ data alone.

It is worth comparing regression models of the form of Equation~\ref{eq:rf_regression} with a potential data fusion approach using multivariate statistical process models, such as those considered in Section~\ref{sec:multivariate}. A multivariate spatio-temporal model could also be used to jointly model dense satellite observations and sparse in situ observations, and then to obtain predictions of the sparsely observed field. In contrast to Equation~\ref{eq:rf_regression}, the spatio-temporal predictor would explicitly depend on the nearby in situ observations and non-colocated satellite observations and would also directly provide an uncertainty estimate. This indicates that multivariate spatio-temporal modeling could hold promise for fusing satellite data with in situ ocean data, but, to the best of our knowledge, the associated computational complexities have so far precluded practical implementations of this approach. Alternatively, one could extend the machine learning regression model to include nearby in situ observations as covariates, which would require using more sophisticated models, such as graph neural networks.

\subsection{Data Assimilation}\label{ss:assimilation}

In this review, we have mostly focused on statistical process models for ocean data. However, we recognize a fundamentally different perspective on ocean modeling through physical dynamical systems. In this perspective, one assumes a model of the form
\begin{equation}
    \bm{x}_{t+1} = M(\bm{x}_t,\bm{p}_t), \label{eq:da_model}
\end{equation}
where $\bm{x}_t$ is the state of the ocean and $\bm{p}_t$ is a collection of control variables at time $t$, and $M$ is a physical model that evolves the state forward in time. In practice, $\bm{x}_t$ is a collection of various oceanographic fields on a discrete grid in longitude, latitude and depth; $\bm{p}_t$ contains fluxes into and out of the ocean (such as surface heat and freshwater fluxes) and other variables (such as surface winds) needed to evolve the state; and $M$ is an ocean general circulation model that solves the Navier--Stokes equations numerically to evolve the state.

Using observations to constrain models of the form of Equation~\ref{eq:da_model} is called \emph{data assimilation} \citep[e.g.,][]{asch2016data}. This approach is particularly powerful for combining observations from heterogeneous instruments in a self-consistent way while respecting physical constraints. Concretely, one has observations of the form $\bm{y}_t = H_t(\bm{x}_t) + \bm{\varepsilon}_t$, where $H_t$ is an observation operator and $\bm{\varepsilon}_t$ represents stochastic noise. The goal is then to use $\bm{y}_t$ to learn $\bm{x}_t$ and often also $\bm{p}_t$. In a rather confusing use of terminology, the resulting estimates are called \emph{reanalysis} data products. There are two broad approaches for data assimilation: Kalman filtering adds stochastic noise to Equation~\ref{eq:da_model} and updates the estimates of $\bm{x}_t$ sequentially while variational methods find the best-fitting $\bm{x}_t$ (and often also $\bm{p}_t$) by solving an optimization problem, which can be understood as finding the Bayesian maximum a posteriori solution to the system in Equation~\ref{eq:da_model}. When a variational estimate is computed simultaneously across all time points, this approach is called 4D-Var data assimilation.

An example of state-of-the-art ocean data assimilation is the ECCO v4 ocean state estimate \citep{forget2015ecco, heimbach2019putting}. ECCO is based on the MITgcm general circulation model \citep{adcroft2004overview} and the 4D-Var variational optimization problem is solved using gradient-based numerical optimization with the gradients computed using the MITgcm adjoint model \citep{heimbach2005efficient}. Each component of the optimization variable has of the order of $10^6-10^7$ dimensions \citep[Table~7]{forget2015ecco} so this is an ultra-high-dimensional optimization problem. ECCO assimilates a wide range of remotely sensed observations, including SST, sea level anomaly and ice cover fraction, with diverse in situ observations ranging from Argo floats and ship-based measurements to observations collected by Southern Ocean elephant seals. The end result is a physically consistent global estimate of key ocean variables and fluxes constrained using a broad range of instruments in the GOOS. Data assimilation has also been used for Lagrangian data and surface drifters, see, for example, \cite{ozgokmen2003assimilation} and \cite{muscarella2015assimilated}.

\section{Conclusion}\label{sec:discandfuture}
In the context of globally deployed oceanic measurement systems, and their associated statistical challenges, we have focused on drifter and float observations. These are a very particular form of spatio-temporal data, either corresponding to tracking a water parcel as it moves horizontally at or near the ocean surface (surface drifters), or vertically profiling the ocean at several depths (profiling floats). Even if we combine these observations, this still corresponds to heavily undersampling the true four-dimensional variability of the ocean. Spatio-temporal statistical models can help plug these observational gaps, as reviewed in this paper, but methodology needs to be carefully chosen to capture the physical quantities of interest to oceanographers, and perform realistically on the available datasets, which are large in size, but sparse in space and time due to the sheer size of the ocean and its many scales of variability. Approaches can vary from traditional statistical approaches, such as stochastic process modeling and spectral analysis, to numerical modeling combined with data assimilation and machine learning for data fusion.

Looking to the future, observational data forms are driven by technological advances. The locational accuracy of tracking, availability of sensor suites, and data transmission technologies have all seen significant advances, promising a future wealth of ocean observations. We anticipate the future rise of low-cost biodegradable technologies, innovating on how the community does experimental design. Such short-lived platforms can also be deployed in the face of a sudden emergency requiring more information in a timely manner, such as an oil spill. An important future experimental design question, where statisticians have a key role, is the \emph{co-design} of ocean observing systems and the associated statistical data analysis methods \cite[e.g.,][]{lascala2023building}. Within this paradigm, the available statistical methods inform observing system design and vice versa, ensuring the optimal combination of observing and data analysis strategies to maximize scientific output.

The rise of machine learning in physical oceanography and climate science \citep{dong2022recent,sinha2021estimating, lyman2023global,eyring2024pushing} shows considerable promise. The limitation of such approaches is that they are predictive rather than explanatory, typically do not provide direct estimates of uncertainty, and their predictive performance needs to be balanced with any reduced interpretability. Connections between machine learning and traditional numerical and statistical methods are particularly interesting, often resulting in more interpretable models. For example, scientific machine learning~\citep{hey2020machine,thiyagalingam2022scientific,cuomo2022scientific} enables augmenting numerical methods with machine learning techniques. 
We anticipate that this will lead to rapid advances in numerical methods, which, when coupled with improved inferential procedures, will enable more ambitious data analysis. 
Machine learning can also be used to define more flexible statistical models, for example, by learning flexible covariance structures in Gaussian process models \citep{wilson2016deep,nag2025modeling}.

The complexity and practical use of many statistical models reviewed in this paper is limited by computational scalability. An attractive use of machine learning in this context is to train fast surrogate models to bypass slow computations. The recent advent of neural simulation-based inference in spatial statistics \citep{gerber2021fast, lenzi2023neural, walchessen2024neural, walchessen2025neural, sainsbury2024likelihood, sainsbury2025neural, sainsbury2025neural2, richards2024neural} enables neural networks to be trained to approximate traditional point estimators, likelihood functions, posteriors, and predictive distributions in spatio-temporal contexts. Once the neural networks are trained, they are usually \emph{amortized} \citep{zammit2025neural} to be efficiently evaluated for new spatial observations without retraining. These techniques hold immense potential in enabling complex, previously intractable, large-scale statistical inference in settings with massive amounts of spatio-temporal data, such as those studied in this article. As such, we expect neural simulation-based inference to be a key driver of future advances in statistical oceanography by enabling the creation of higher-resolution data products, the use of higher-fidelity statistical models and scalable multivariate inference, including joint modeling of in situ data with high-resolution satellite observations.

\section*{Acknowledgments}
The authors are grateful to Dr Jeffrey Early at NorthWest Research Associates, Seattle, USA, for helpful discussions and feedback.


\end{document}